\documentclass[sigconf]{acmart} 

\usepackage{subfigure}
\usepackage{amsmath,amsfonts}
\usepackage[algo2e,linesnumbered,ruled,lined,noend]{algorithm2e}
\SetKwInOut{Parameter}{Parameters}
\SetKwInOut{Constraint}{Constraints}
\SetKwFunction{Mining}{Mining}
\SetKwFunction{MiningTwo}{Mining2}
\SetKwFunction{Candidates}{Candidates}
\SetKwProg{Fn}{Function}{}{}
\usepackage{graphicx}
\usepackage{textcomp}
\usepackage{color}
\usepackage{xcolor}
\usepackage{xspace}
\def\BibTeX{{\rm B\kern-.05em{\sc i\kern-.025em b}\kern-.08em
    T\kern-.1667em\lower.7ex\hbox{E}\kern-.125emX}}
    
\usepackage{multirow} 
\usepackage{booktabs} 

\usepackage[normalem]{ulem}
\useunder{\uline}{\ul}{}
\usepackage[table]{xcolor}

\usepackage{pgf}
\usepackage{enumitem}
\setlist[itemize]{leftmargin=*}
\usepackage{stfloats}

\usepackage{makecell}

\usepackage{bbold}
\usepackage{mathtools}
\usepackage{comment}
\usepackage{microtype}
\usepackage[misc]{ifsym}
\usepackage{url}
\SetCommentSty{mycommfont}

\newcommand{\method}{UF\textsuperscript{2}O\xspace}
\newcommand{\std}[1]{{\small (#1)}}
\newcommand{\best}{\cellcolor{green!70}}
\newcommand{\bestbox}{%
  {\setlength{\fboxsep}{0.6pt}%
  \colorbox{green!70}{green}}%
}

\AtBeginDocument{%
  \providecommand\BibTeX{{%
    \normalfont B\kern-0.5em{\scshape i\kern-0.25em b}\kern-0.8em\TeX}}}

\copyrightyear{2027}
\acmYear{2027}
\setcopyright{cc}
\setcctype{by}
\acmConference[WSDM '27]{Proceedings of the Twentieth ACM International Conference on Web Search and Data Mining}{February 15--19, 2027}{Hong Kong, Hong Kong}
\acmBooktitle{Proceedings of the Twentieth ACM International Conference on Web Search and Data Mining (WSDM '27), February 15--19, 2027, Hong Kong, Hong Kong}
\acmDOI{10.1145/3862322.3863490}
\acmISBN{979-8-4007-3117-4/2027/02}



\begin{document}

    \title[User-Adaptive Fusion of Multiple Foundation Models for Recommender Systems]{User-Adaptive Fusion of Multiple Foundation Models \\ for Recommender Systems}
    
    \settopmatter{authorsperrow=3}
	
    \author{Sunwoo Kim}
	\affiliation{%
    	\institution{KAIST}
            \city{Seoul}
            \country{Republic of Korea}
	}
	\email{kswoo97@kaist.ac.kr}

    \author{Hyunjin Hwang}
	\affiliation{%
    	\institution{KAIST}
            \city{Seoul}
            \country{Republic of Korea}
	}
	\email{hyunjinhwang@kaist.ac.kr}

        \author{Geon Lee}
	\affiliation{%
		\institution{KAIST}
            \city{Seoul}
            \country{Republic of Korea}
	}
	\email{geonlee0325@kaist.ac.kr}

    \author{Kyungho Kim}
	\affiliation{%
		\institution{KAIST}
            \city{Seoul}
            \country{Republic of Korea}
	}
	\email{kkyungho@kaist.ac.kr}
	
	\author{Kijung Shin}
	\affiliation{%
		\institution{KAIST}
            \city{Seoul}
            \country{Republic of Korea}
	}
	\email{kijungs@kaist.ac.kr}
	
    \begin{abstract}

\textit{Can recommender systems benefit from combining a collection of foundation models within and across modalities?}
While different foundation models may capture complementary aspects of item metadata, we find that naively combining their representations yields only marginal gains over using a single foundation model. 
We make a key observation: \textit{the users who benefit vary across foundation models.}
Motivated by this observation, we introduce \method, a user-adaptive approach for fusing multiple foundation models for recommendation. 
In a nutshell, \method models the contribution of each foundation model as user-specific by maintaining separate recommendation views and adaptively combining their scores for each user.
By performing fusion at the score level, \method can also fuse heterogeneous foundation models without directly aligning their representation spaces.
Extensive experiments that jointly combine four foundation models on four datasets show that \method outperforms existing fusion approaches in both static and sequential recommendation settings, with improvements of up to 7\% over the strongest fusion baseline.
The code, datasets, and supplementary materials for our work are available at~\url{https://github.com/kswoo97/UFFO}.

    \end{abstract}
	
	\maketitle

    \section{Introduction}
    \label{sec:intro}
    
Recommender systems are central to various web services, retrieving items that align with user interests from extensive item catalogs~\cite{liu2026hybrid, kim2026memorization, huang2025large, lee2026rethinking, xu2026mmq, kim2026retrieval, kim2025self, kim2026itemrag, kim2024towards, kim2025multi}. 
By inferring user preferences from historical item interactions such as purchases, recommender systems deliver personalized recommendations that enhance user satisfaction and contribute to business revenue.

A variety of approaches leverage foundation models’ ability to capture semantic information from item metadata for recommendation~\cite{liu2024llm, hu2025alphafuse, luo2026time, kim2026personalized, kim2026training}. 
Among them, one promising direction is to leverage foundation models to encode item metadata (e.g., titles and images), and use the resulting embeddings as item representations in recommender-system backbones such as LightGCN~\cite{he2020lightgcn} or SASRec~\cite{kang2018self}, a strategy widely explored in many studies~\cite{li2023text, zhang2024dual, liu2025llmemb, sheng2025language}.

\begin{figure}[t]
    \centering
    \includegraphics[width=\linewidth]{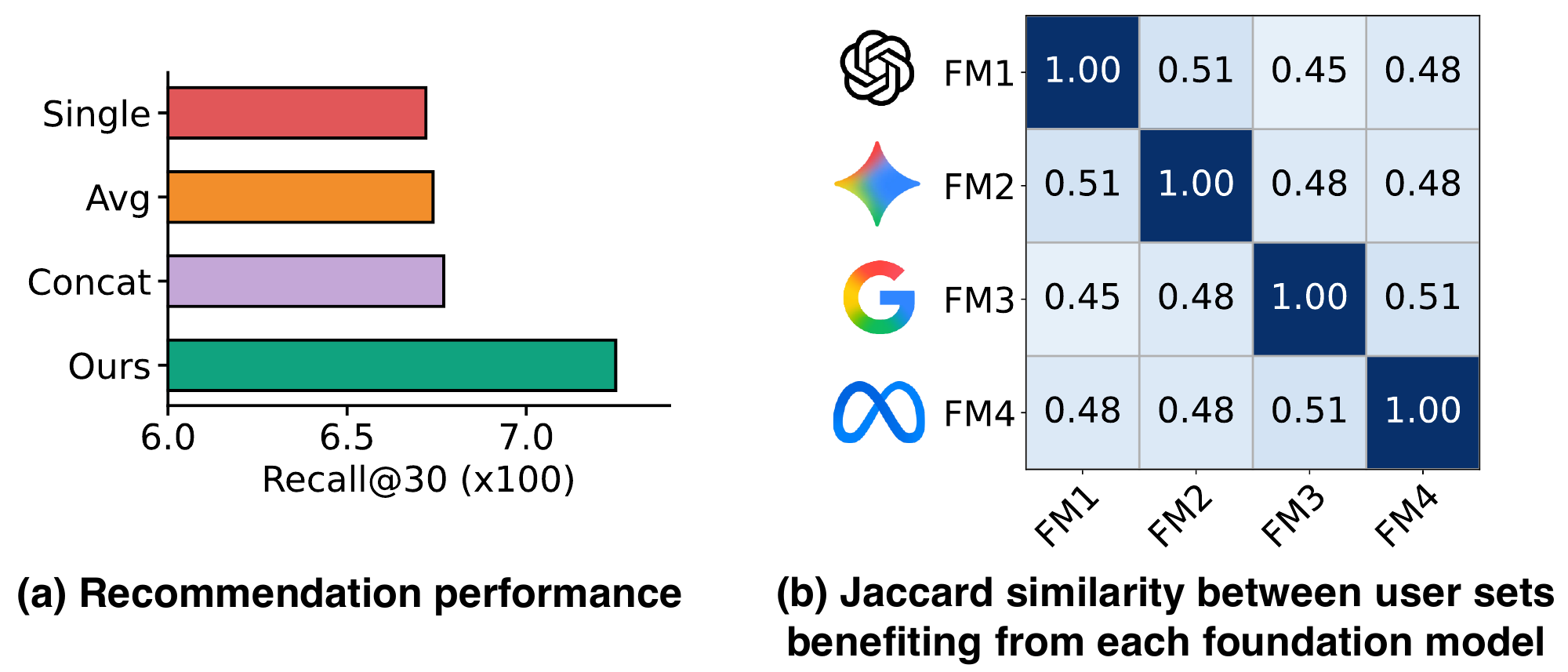}
    \caption{\textit{Key motivation.}
    Compared with using a single foundation model (\textit{Single}), straightforward fusion of embeddings from multiple foundation models, such as averaging (\textit{Avg}) or concatenation (\textit{Concat}), provides only marginal performance improvements, as shown in (a).
    Our analysis reveals that different foundation models vary in the users they benefit. Specifically, as reported in (b), the sets of users whose next purchases appear in the Top-$K$ recommendations based on each foundation model vary across models. 
    This observation motivates the design of \method, which fuses foundation models in a user-adaptive way.
    As shown in (a),
    \method (\textit{Ours}) substantially outperforms the simple fusion baselines.
    Results are from the Sports dataset under the sequential setting.}
    \label{fig:motivation}

\end{figure}

Prior work in natural language processing and computer vision suggests that different foundation models, even within the same modality, may capture different aspects of the same input data~\cite{garcia2021benchmarking, zhang2024unveiling, kiela2018dynamic}. 
For example, given the same input image, some vision foundation models tend to focus more on texture, whereas others tend to capture the overall shape and spatial organization of objects~\cite{ramtoula2026fantastic}.
This suggests that the information extracted from item metadata can vary depending on the foundation model used for encoding, even within the same modality, yet this property has received limited attention in the recommendation literature.


Motivated by this characteristic, we \textit{explore the use of a collection of foundation models} to encode item metadata for recommendation. 
However, simply incorporating multiple foundation models does not necessarily lead to meaningful improvements in recommendation performance.
As shown in Figure~\ref{fig:motivation}(a), simply concatenating or averaging item embeddings derived from different foundation models yields little or no improvement over using a single foundation model. 
From this observation, we hypothesize that effectively leveraging multiple foundation models requires a more sophisticated fusion strategy that preserves model-specific representations while adaptively determining their contributions to recommendation.

In light of this challenge, we propose \textbf{\method} (\textbf{\underline{U}}ser-adaptive \textbf{\underline{F}}usion of \textbf{\underline{FO}}undation models), a method for user-adaptive score-level fusion of multiple foundation models for recommendation.
Its design is motivated by our empirical analysis, which suggests that the users who benefit vary across foundation models.
Here, we say that a foundation model benefits a user if the user’s test item appears in the Top-$K$ recommendations generated by a recommender system using item representations from that foundation model.
As shown in Figure~\ref{fig:motivation}(b), the sets of users benefiting from different foundation models overlap but are not identical, with an average Jaccard similarity of 0.48. 
This observation motivates modeling the contribution of each foundation model on a per-user basis, rather than applying the same fusion strategy to all users.

Building on this finding, \method adopts a user-adaptive score-level fusion approach. 
Rather than directly combining heterogeneous embeddings from different foundation models into a single item representation shared across users, \method preserves a separate recommendation view for each model and performs fusion only after obtaining their recommendation scores.
This design allows foundation models with different architectures, modalities, and representation spaces to contribute without requiring direct alignment of their embeddings or risking the loss of model-specific information.

Specifically, for each user–item pair, \method first constructs a separate recommendation view for each foundation model by using its item representations as input to the recommender-system backbone. 
It then independently computes a recommendation score from each view and combines these scores through a user-conditioned attention module, which assigns different importance to each foundation model according to the target user.
This module assigns user-specific weights to the recommendation scores and computes their weighted average, thereby increasing or decreasing the contribution of each foundation model to the final recommendation.  
Moreover, we introduce a regularization technique that helps \method effectively leverage multiple foundation models by preventing the fusion weights from collapsing onto a single model.
Notably, \method is applicable to both static and sequential recommendation settings.

Through extensive experiments on 4 datasets with 13 baseline methods in each recommendation setting, we demonstrate the effectiveness of \method for recommendation. 
Notably, \method jointly fuses four foundation models and outperforms existing foundation-model fusion baselines, achieving up to a 7\% improvement over the strongest fusion baseline.
Moreover, \method outperforms several multimodal recommender systems in 15 out of 16 settings, highlighting the effectiveness of leveraging multiple foundation models, including multiple models within each modality.
Remarkably, despite using multiple foundation models, \method incurs less than $10^{-4}$ seconds of additional inference time per user compared with using a single foundation model. 
This minimal overhead is enabled by pre-computing item embeddings and processing multiple foundation-model views in parallel during inference.
Our key contributions are as follows:

\begin{itemize}
    \item \textbf{\textit{C1. New observations.}} We show that foundation models vary in the users for whom they provide accurate recommendations.
    \item \textbf{\textit{C2. New method.}} We propose \method, a user-adaptive score-level approach for fusing multiple foundation models, together with a regularization technique for stable training.
    \item \textbf{\textit{C3. Strong performance.}} By adaptively fusing four foundation models, \method outperforms existing fusion approaches in both sequential and static recommendation settings.
\end{itemize}

The code, datasets, and supplementary materials for our work are available at~\url{https://github.com/kswoo97/UFFO}.


    \section{Related work and preliminaries}
    \label{sec:relatedwork}
    

\subsection{Related work}\label{subsec:relatedwork}

\subsubsection{\textbf{Combining multiple foundation models.}}
Studies in natural language processing (NLP) and computer vision (CV) have shown that different foundation models may capture different aspects of the same item metadata~\cite{garcia2021benchmarking, zhang2024unveiling, ramtoula2026fantastic, kiela2018dynamic}.
Prior work has attributed this tendency to differences in foundation model architectures and the training datasets used by each model~\cite{peters2018dissecting, ramtoula2026fantastic}.
Motivated by this characteristic, many approaches combine \textit{multiple representations} obtained from diverse foundation models to construct a richer representation that captures diverse aspects of the data~\cite{kiela2018dynamic, shen2024mome, dong2023simmmdg}.
These approaches typically learn a common fused representation shared across users, implicitly treating the utility of each foundation model as user-independent.
In contrast, \method treats foundation-model utility as user-dependent, preserves a separate recommendation view for each model, and performs user-adaptive fusion at the score level.

\subsubsection{\textbf{Item encoding using a foundation model.}}
Various recommender systems leverage item metadata, such as titles, descriptions, and images, to capture rich semantic information about items~\cite{yue2025preference, xie2026joint, li2025teach, chen2026factorized, v2025contrastive}.
Among these approaches, a promising direction is to encode item metadata using \textit{pretrained foundation models} and incorporate the resulting embeddings to obtain the item representations~\cite{liu2024once, hou2023learning, yuan2023go, kim2026personalized}.
Specifically, these embeddings can either (1) be used directly as item representations~\cite{yuan2023go} or (2) be combined with transductive item ID embeddings~\cite{kim2026personalized}.
One of the research areas most closely related to our study is multimodal recommendation~\cite{ong2025spectrum, chen2026modalities, chen2025hypercomplex, ren2025modality}, which leverages multimodal representations of item metadata for recommendation. 
Our setting is complementary to conventional multimodal recommendation. While multimodal recommenders primarily study how to combine different modalities, they typically rely on a single encoder for each modality. 
We instead study how to exploit multiple independently pretrained foundation models, including multiple models within the same modality, and therefore face the additional challenge of combining heterogeneous model-specific representations.

\subsection{Preliminaries}\label{subsec:preliminary}

\subsubsection{\textbf{Item information.}}
Let $\mathcal{I}$ denote the item set.
Each item $i_k \in \mathcal{I}$ is associated with its corresponding metadata $\mathcal{D}_{k}$.
Such metadata includes textual information (e.g., title or description) and/or visual information (e.g., image).
We denote the set of foundation-model-based encoders by $\mathcal{F}=\{f^{(1)},f^{(2)},\cdots,f^{(\vert \mathcal{F}\vert)}\}$.
Lastly, we denote the metadata embedding of item $i_{k}$ produced by a foundation model $f^{(p)}$ as $\mathbf{x}^{(p)}_{k}$ (i.e., $\mathbf{x}^{(p)}_{k} = f^{(p)}(\mathcal{D}_k)$).
Note that a foundation model need not jointly process multiple modalities; it may operate on a single modality. For example, Gemini Text Embedding~\cite{lee2025gemini} encodes only textual information.

\subsubsection{\textbf{Static setting.}}
Let $\mathcal{U}$ denote the user set.
In the static setting, we assume that a user-item bipartite graph $\mathcal{G} = (\mathcal{U} \cup \mathcal{I},\mathcal{E})$ is given, where $\mathcal{E}$ denotes the set of observed user-item interactions.
Specifically, an interaction between a user $u \in \mathcal{U}$ and an item $i_{k} \in \mathcal{I}$ is represented by an edge $(u,i_{k}) \in \mathcal{E}$.

\subsubsection{\textbf{Sequential setting.}}
As in the static setting, we let $\mathcal{U}$ denote the user set.
In the sequential setting, each user $u \in \mathcal{U}$ is associated with an interaction sequence $s_u = (i_{u,1}, i_{u,2}, \ldots, i_{u,n_u})$, where $i_{u,t} \in \mathcal{I}$ denotes the $t$-th item with which user $u$ interacted in chronological order, and $n_u$ denotes the sequence length.
    
    \section{Analysis}
    \label{sec:analysis}

\begin{figure*}[t]
    \centering
    \includegraphics[width=.85\textwidth]{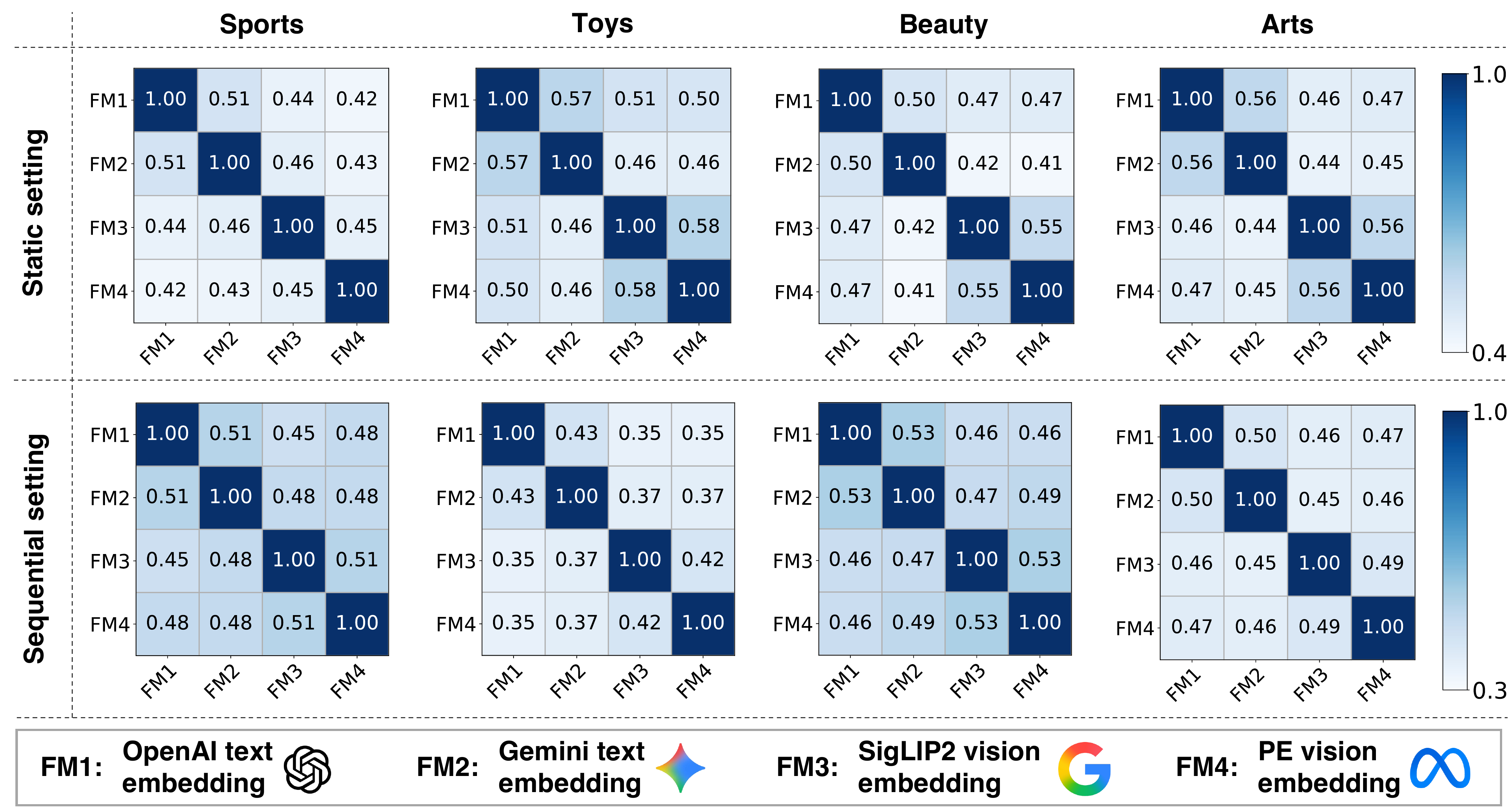}
    \caption{\textit{Overlap of users benefiting from different foundation models.}
    For each pair of foundation models, the Jaccard similarity between the set of users who receive accurate recommendations (i.e., users whose next purchases appear in the Top-$K$ recommendations)
    from one foundation model and the corresponding set for the other foundation model is reported.
    The average Jaccard similarity is below 0.5 in both static and sequential recommendation settings, with values of 0.48 and 0.46, respectively.}
    \label{fig:mainanalysis}
\end{figure*}

In this section, we present an analysis showing that foundation models vary in the users they benefit, providing the key motivation for our proposed method.
To this end, we first describe the analysis setup (Section~\ref{subsec:analysis-setup}), followed by the analysis results and their key implications (Section~\ref{subsec:analysis-result}).

\subsection{Analysis setting}\label{subsec:analysis-setup}

\subsubsection{\textbf{Datasets.}}
We use four domain-specific e-commerce datasets released by~\citet{hou2024bridging}: (1) Sports \& Outdoors (Sports), (2) Toys \& Games (Toys), (3) Beauty \& Personal Care (Beauty), and (4) Arts, Crafts, and Sewing (Arts). 
We use the preprocessed versions of these datasets provided by~\citet{kim2026personalized}.
Here, we use textual and visual metadata, with each item’s title serving as its textual metadata and its image serving as its visual metadata.

\subsubsection{\textbf{Foundation models.}}
We employ four foundation models to encode item metadata: two text foundation models and two vision foundation models.
The two text foundation models are OpenAI’s text-embedding-3-large~\cite{openai_text_embedding_3_large} and Gemini Text Embedding~\cite{lee2025gemini}, both of which take each item’s title as input.
The two vision foundation models are SigLIP2~\cite{tschannen2025siglip} and Perception Encoder (PE)~\cite{bolya2026perception}, both of which take each item's image as input.
In the following analysis, we consider each foundation model separately and compare the resulting recommendations.

\subsubsection{\textbf{Forward pass and training in the static setting.}}\label{subsec:analysis-static-setup}
In the static recommendation setting, we use LightGCN~\cite{he2020lightgcn} as the backbone recommender system.
We use learnable ID-based transductive embeddings for each user $u \in \mathcal{U}$, denoted by $\mathbf{t}_{u}\in\mathbb{R}^{d}$, and each item $i_{k} \in \mathcal{I}$, denoted by $\textbf{t}_{k} \in \mathbb{R}^{d}$.

Consider using a single foundation model $f^{(p)}$.
For each item $i_{k}\in\mathcal{I}$, we encode its metadata using $f^{(p)}$ to obtain its metadata embedding $\mathbf{x}^{(p)}_{k}=f^{(p)}(\mathcal{D}_k)$.
We then use an MLP projector, denoted by $\texttt{MLP}^{(p)}$, to map the item's metadata embedding $\mathbf{x}^{(p)}_{k}$ to a $d$-dimensional embedding $\mathbf{\tilde{x}}^{(p)}_{k} \in \mathbb{R}^{d}$ (i.e., $\mathbf{\tilde{x}}^{(p)}_{k}=\texttt{MLP}^{(p)}(\mathbf{x}^{(p)}_{k})$), matching the dimensionality of the user and item ID embeddings.

We then construct two views: an ID view and a metadata view.
The two views share the user ID embedding $\mathbf{t}_{u}$ as the input user representation for user $u$, while using different item representations.
Specifically, the ID view uses the ID embedding $\mathbf{t}_{k}$ for item $i_{k}$, whereas the metadata view uses the projected metadata embedding $\mathbf{\tilde{x}}^{(p)}_{k}$.
Using these input representations, we independently perform LightGCN-based message passing over the user–item interaction graph $\mathcal{G}$ for the two views, obtaining view-specific final user and item representations. 

Finally, for a given user-item pair, we compute a view-specific recommendation score as the inner product of the final user and item representations in that view, and sum the scores from the two views to obtain the final recommendation score.
All trainable parameters, including the user and item ID embeddings and the MLP projector, are optimized using the BPR loss~\cite{rendle2009bpr}, following the training procedure adopted in the original LightGCN paper~\cite{he2020lightgcn}.

\subsubsection{\textbf{Forward pass and training in the sequential setting.}}\label{subsec:analysis-sequential-setup}
In the sequential recommendation setting, we use SASRec~\cite{kang2018self} as the backbone recommender system.
Unlike in the static setting, we use only item ID embeddings ($\mathbf{t}_{k}$ for each $i_k \in \mathcal{I}$) and omit user ID embeddings, following standard practice in sequential recommendation~\cite{kang2018self, zhou2020s3, sun2019bert4rec}.
For item $i_k$, we obtain the projected metadata embedding $\mathbf{\tilde{x}}^{(p)}_{k}$ from the single foundation model $f^{(p)}$ and the projector $\texttt{MLP}^{(p)}$ as in the static setting. 

Similar to the static setting, we construct two views: an ID view and a metadata view. The ID view uses the transductive item embeddings $\mathbf{t}_{k}=\mathbf{\tilde{x}}^{(ID)}_{k}, \forall i_k \in \mathcal{I}$ as item representations, whereas the metadata view uses $\mathbf{\tilde{x}}^{(p)}_{k}, \forall i_k \in \mathcal{I}$ as item representations.
Then, for a user $u$ with an interaction sequence $s_u = (i_{u,1}, i_{u,2}, \ldots, i_{u,n_u})$, we feed the representation sequence $[\mathbf{\tilde{x}}^{(q)}_{u,1}, \mathbf{\tilde{x}}^{(q)}_{u,2}, \ldots, \mathbf{\tilde{x}}^{(q)}_{u,n_u}]$ for each $q \in \{p,\mathrm{ID}\}$ into SASRec, where $\mathbf{\tilde{x}}^{(q)}_{u,t}$ denotes the representation of item $i_{u,t}$ in view $q$.
To compute the recommendation score of item $i_j$ as the next item for user $u$, we take the inner product between the SASRec output at the $n_u$-th position and the item representation $\mathbf{\tilde{x}}^{(q)}_{j}$ for each $q \in \{p,\mathrm{ID}\}$. 
We then sum the scores from the two views to obtain the final score.
All trainable parameters, including the item ID embeddings, the MLP projector, and SASRec parameters, are optimized with the binary-cross-entropy-based training procedure adopted in the original SASRec paper~\cite{kang2018self}.

\subsubsection{\textbf{Final evaluation protocol.}}\label{subsec:analysis-user}
After training, we measure the extent to which the sets of users who benefit from different foundation models overlap.
Here, we say that a foundation model benefits a user if the user’s next purchase (i.e., test item) appears in the Top-$K$ recommendations produced by a recommender system that uses item representations derived from that foundation model.
We conduct this analysis in both sequential and static settings.
Specifically, for each foundation model $f^{(p)} \in \mathcal{F}$, we first train recommender systems in the static and sequential settings following the procedures described in Sections~\ref{subsec:analysis-static-setup} and~\ref{subsec:analysis-sequential-setup}, respectively.
After training, for each foundation model $f^{(p)}$, we obtain a top-30 recommendation list for every user and define $\mathcal{U}^{(p)} \subseteq \mathcal{U}$ as the set of users whose next purchase appears in the corresponding recommendation list.
Lastly, for every foundation model pair $\{f^{(p)},f^{(q)}\} \in \binom{\mathcal{F}}{2}$, we measure the Jaccard similarity between $\mathcal{U}^{(p)}$ and $\mathcal{U}^{(q)}$ (i.e., $(\vert \mathcal{U}^{(p)}\cap \mathcal{U}^{(q)} \vert) / (\vert \mathcal{U}^{(p)} \cup \mathcal{U}^{(q)} \vert)$).

\subsection{Analysis results and implications}\label{subsec:analysis-result}
Analysis results are presented in Figure~\ref{fig:mainanalysis}.
Overall, the sets of users benefiting from different foundation models overlap but are not identical, with average Jaccard similarities of 0.48 and 0.46 in the static and sequential recommendation settings, respectively.
In addition, foundation models using different input modalities exhibit lower overlap than those using the same modality, with average Jaccard similarities of 0.45 and 0.51, respectively.
For comparison, when we compare the same foundation model trained with different random seeds, the average Jaccard similarity is around 0.8, substantially higher than the similarity observed across different foundation models; details are in our supplementary material~\cite{anonymousgithub}.
These results show that different foundation models benefit different users, 
suggesting that applying the same fusion strategy to all users may not fully exploit their distinct recommendation signals. 
Together with the limited effectiveness of simple representation-level fusion observed in Figure~\ref{fig:motivation}, this motivates
preserving the distinct signals of individual foundation models and adaptively combining them for each user.


\begin{figure*}[t]
    \centering
    \includegraphics[width=0.85\textwidth]{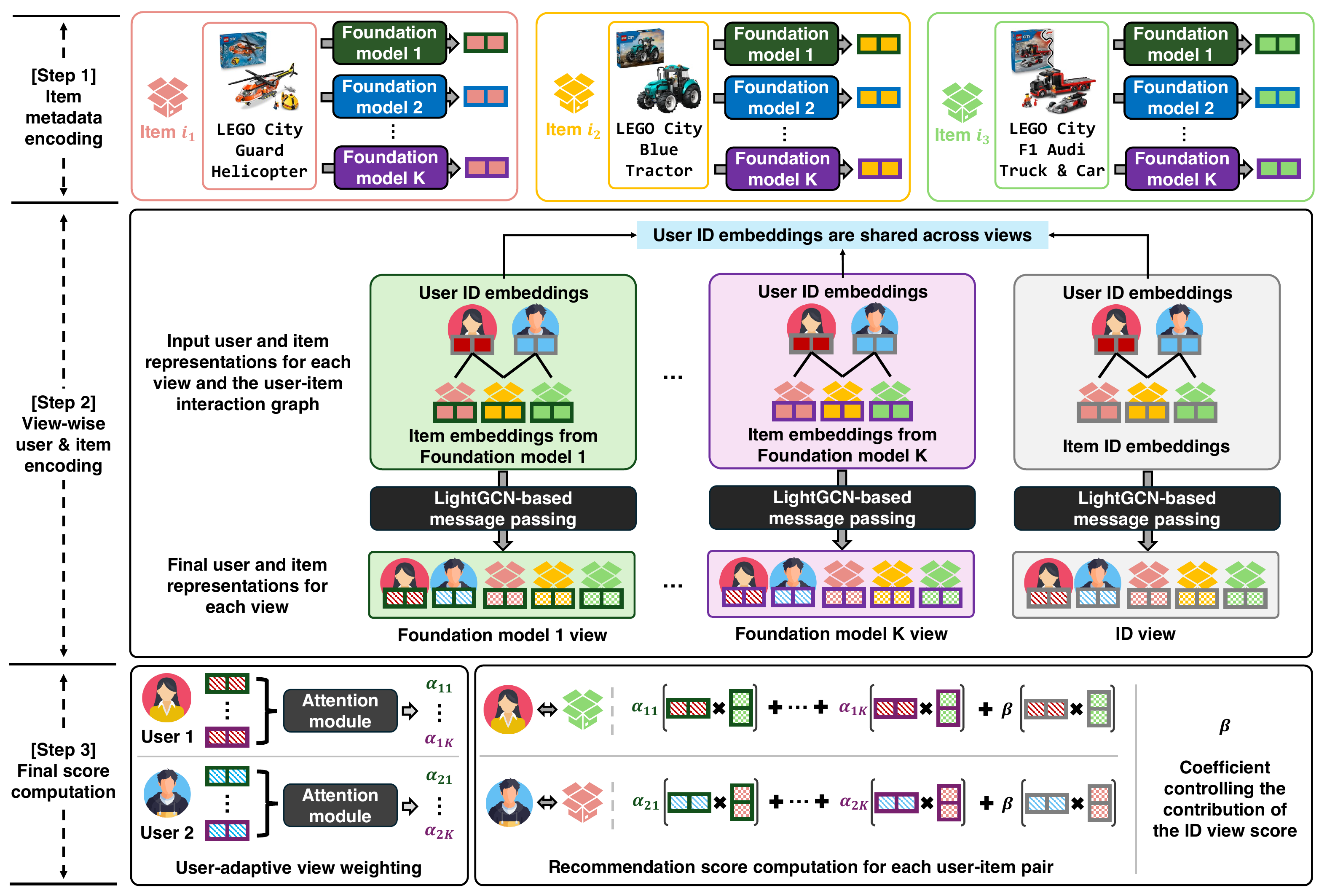}
    \vspace{-2mm}
    \caption{\textit{Overview of \method under the static setting.}
    The forward-pass process of \method consists of three steps.
    In the first step, \method encodes the metadata of each item using multiple foundation models (see the first row).
    In the second step, \method separately obtains user and item representations for each view, using LightGCN~\cite{he2020lightgcn} as the backbone recommender in the static setting and SASRec~\cite{kang2018self} in the sequential setting (see the second row).
    Lastly, in the third step, \method computes a weighted average of the scores from different foundation models using a user-adaptive weighting strategy (see the third row).
    }
    \label{fig:mainmethod}
\end{figure*}

    \section{Proposed method: \method}
    \label{sec:method}

We introduce \textbf{\method} (\textbf{\underline{U}}ser-adaptive \textbf{\underline{F}}usion of \textbf{\underline{FO}}undation models), a user-adaptive approach for fusing multiple foundation models in recommender systems.
Motivated by the observations in Sections~\ref{sec:intro} and~\ref{sec:analysis}, \method follows two key design principles: (1) preserving foundation-model-specific recommendation signals rather than directly merging heterogeneous representations into a single representation, and 
(2) adaptively determining their contributions according to the target user.
To achieve this, \method constructs a separate recommendation view for each foundation model by using its item representations as input to the backbone recommender system, with each view independently producing a recommendation score.
\method then performs user-adaptive score-level fusion across these views through a user-conditioned attention module.
An overview of \method in the static setting is presented in Figure~\ref{fig:mainmethod}, and its sequential counterpart can be found in the supplementary material~\cite{anonymousgithub}.
Below, we first describe how item metadata are encoded using foundation models (Section~\ref{subsec:method-dataencoding}). 
We then describe how foundation-model-specific recommendation scores are computed (Section~\ref{subsec:method-scoring}) and subsequently fused in a user-adaptive manner (Section~\ref{subsec:method-fusion}).
After that, we describe our training strategy with a specialized regularization loss and the inference-time computation of final recommendation scores (Section~\ref{subsec:method-training}).

\subsection{Item metadata encoding}\label{subsec:method-dataencoding}
The item metadata encoding process of \method is identical in both static and sequential settings. 
For the encoding, we use multiple foundation models to obtain multiple metadata embeddings for each item.
Consider an item $i_k$ and a set of foundation models $\mathcal{F}=\{f^{(1)}, f^{(2)}, \cdots ,f^{(\vert \mathcal{F}\vert)}\}$.
For each foundation model $f^{(p)} \in \mathcal{F}$, we encode the metadata $\mathcal{D}_k$ of item $i_k$ to obtain its foundation-model-specific representation $\mathbf{x}^{(p)}_k$ (i.e., $\mathbf{x}^{(p)}_k = f^{(p)}(\mathcal{D}_k)$).
Note that all foundation models remain frozen and are not updated end-to-end using the recommendation loss, as we aim to preserve each model’s inherent ability to capture foundation-model-specific aspects of items.
Moreover, (1) the weights of many foundation models are not publicly available, and (2) fine-tuning multiple foundation models would incur substantial computational costs.
After encoding the metadata, we project each resulting embedding into a $d$-dimensional feature space to ensure dimensional compatibility with the user and/or item ID embeddings.
Specifically, for each foundation-model-specific representation $\mathbf{x}^{(p)}_k$, where $p \in [\vert\mathcal{F}\vert]$, we use a foundation-model-specific projector $\texttt{MLP}^{(p)}$ to obtain the projected representation $\mathbf{\tilde{x}}^{(p)}_k \in \mathbb{R}^{d}$ (i.e., $\mathbf{\tilde{x}}^{(p)}_k = \texttt{MLP}^{(p)}(\mathbf{x}^{(p)}_k)$).
Note that all parameters of the MLP projectors $\texttt{MLP}^{(p)}, \forall p \in [|\mathcal{F}|]$, are learnable and trained end-to-end using the recommendation loss.

\subsection{Foundation-model-specific scoring}\label{subsec:method-scoring}
After obtaining item metadata representations from each foundation model, we compute a separate recommendation score for each model rather than first collapsing their representations into a single embedding.
This converts heterogeneous representations from different foundation models into user-item recommendation scores, providing a common basis for combining them.
Specifically, we treat each foundation model as a distinct view that leverages the item representations generated by the corresponding model.
Since the score computation process differs between the static and sequential settings, we describe each process separately below.

\subsubsection{\textbf{Score computation in the static setting.}}
We also include an ID view that uses transductive item ID embeddings, alongside the $\vert\mathcal{F}\vert$ foundation-model views.
For the backbone recommender system in the static setting, \method uses LightGCN~\cite{he2020lightgcn}, as adopted in Section~\ref{subsec:analysis-static-setup}.
Specifically, for the view associated with foundation model $f^{(p)}$, we use the projected representations $\mathbf{\tilde{x}}^{(p)}_k \in \mathbb{R}^d$ of items $i_k \in \mathcal{I}$ as the input item representations.
For the ID view, we use transductive item ID embeddings as the input item representations.
Across all views, we share the same input user representations, given by transductive user ID embeddings $\mathbf{t}_{u}$ for all $u \in \mathcal{U}$.

We then independently perform LightGCN-based message passing for each view.
Consequently, for each view $p \in [|\mathcal{F}|] \cup \{{ID}\}$, we obtain the final user representation $\mathbf{e}^{(p)}_{u} \in \mathbb{R}^d$ for each $u\in\mathcal{U}$ and item representation $\mathbf{e}^{(p)}_{k} \in \mathbb{R}^d$ for each item $i_k \in \mathcal{I}$.
Note that, although different views share the same input user embeddings, their final user representations differ because message passing is based on view-specific item representations.
Lastly, to compute the recommendation score of item $i_k$ for user $u$, we obtain a view-specific score for each view $p$ as $s^{(p)}_{uk} = (\mathbf{e}^{(p)}_{u})^{T}\mathbf{e}^{(p)}_{k} \in \mathbb{R}$.
In other words, we obtain $\vert\mathcal{F}\vert + 1$ scores for each user-item pair.

\subsubsection{\textbf{Score computation in the sequential setting.}}
In the sequential setting, we also include an ID view based on transductive item ID embeddings alongside the $\vert\mathcal{F}\vert$ foundation-model views.
In this setting, for the backbone recommender system, \method uses SASRec~\cite{kang2018self}, with its parameters shared across views.
Specifically, for the view associated with foundation model $f^{(p)}$, we use the projected metadata representations $\mathbf{\tilde{x}}^{(p)}_k \in \mathbb{R}^d$ of all items $i_k \in \mathcal{I}$ as the item representations.
For the ID view, we use transductive item embeddings $\mathbf{t}_{k}\in \mathbb{R}^{d}$ of items $i_k \in \mathcal{I}$ as the item representations.

Then, for user $u$ with an interaction sequence $s_u = (i_{u,1}, i_{u,2}, \ldots,$ $i_{u,n_u})$, 
we feed the representation sequence $[\mathbf{\tilde{x}}^{(p)}_{u,1},\mathbf{\tilde{x}}^{(p)}_{u,2}, \ldots ,\mathbf{\tilde{x}}^{(p)}_{u,n_u}]$ into SASRec, where $\mathbf{\tilde{x}}^{(p)}_{u,t}$ denotes the item representation of item $i_{u,t}$, for each view $p \in [\vert\mathcal{F}\vert] \cup \{ID\}$.
To compute the recommendation score for the next item that user $u$ is likely to interact with, we use the SASRec output embedding corresponding to the user’s last interacted item $i_{u,n_u}$, denoted as $\mathbf{e}_{u}^{(p)} \in \mathbb{R}^{d}$, as the final user representation.
Specifically, to compute the recommendation score of item $i_k$ as the next interaction for user $u$, we obtain a view-specific score for each view $p \in [\vert\mathcal{F}\vert] \cup \{ID\}$ as $s^{(p)}_{uk} = (\mathbf{e}^{(p)}_{u})^{T}\mathbf{\tilde{x}}^{(p)}_{k} \in \mathbb{R}$.

\subsection{User-adaptive score fusion}\label{subsec:method-fusion}

Recall that a key observation from Section~\ref{sec:analysis} is that the users who benefit vary across foundation models.
Therefore, in \method, we adapt the contribution of each foundation model to the target user when computing the final recommendation score.
To this end, we employ a user-adaptive attention mechanism that assigns user-specific weights to the recommendation scores produced by each foundation-model view and computes their weighted average to obtain the final recommendation score.
Consider computing the recommendation score of item $i_k$ for user $u$.
First, we use a learnable projection vector $\mathbf{a}\in \mathbb{R}^{d}$ to project each final user representation $\mathbf{e}^{(p)}_u$ to a scalar attention logit.
We then apply Softmax to these logits across the views to obtain the corresponding attention {weights}:
\begin{equation}\label{eq:attention}
    [\alpha_{u1},\alpha_{u2},\cdots,\alpha_{u\vert \mathcal{F}\vert}]=
    \texttt{Softmax}(\frac{\mathbf{a}^{T}\mathbf{e}^{(1)}_{u}}{\tau},
    \frac{\mathbf{a}^{T}\mathbf{e}^{(2)}_{u}}{\tau},
    \cdots,
    \frac{\mathbf{a}^{T}\mathbf{e}^{(\vert\mathcal{F}\vert)}_{u}}{\tau}),
\end{equation}
where $\alpha_{up} \in [0,1]$ denotes the attention weight assigned to foundation-model view $p$ for user $u$, 
and $\tau$ denotes the temperature hyperparameter.
Unlike globally shared fusion weights, $\alpha_{up}$ assigns user-specific contributions to each foundation model, reflecting the user-level heterogeneity observed in Section~\ref{sec:analysis}.
Then, based on these attention weights, {we compute a weighted average of the foundation-model scores.}
In both settings, the recommendation score of item $i_k$ for user $u$, denoted by $s_{uk}$, is computed as follows:
$s_{uk} = \sum^{\vert\mathcal{F}\vert}_{p=1}\alpha_{up}s^{(p)}_{uk} + \beta s_{uk}^{({ID})}$, where $\beta$ is a hyperparameter that controls the contribution of the ID-view score relative to the foundation-model-view scores.
Note that we empirically observe comparable score magnitudes across foundation-model views (see~\cite{anonymousgithub}), {suggesting that no single view dominates the final score solely due to its score scale.\footnote{{We can explicitly align the scales of scores across views through standardization, but doing so reduces recommendation performance in our experiments (see~\cite{anonymousgithub}).}}
Accordingly, the attention weights control each view’s relative contribution, with higher weights giving greater influence to that view’s predictions.}

\subsection{Training and inference}\label{subsec:method-training}

\subsubsection{Training}
The training loss of \method consists of two components: 
(1) a recommendation supervision loss and 
(2) an attention-collapse regularization loss that prevents the attention {weights} from collapsing onto a single foundation-model view.

For the recommendation supervision loss $\mathcal{L}_{rec}$, we adopt the training objective used in the original backbone recommender:
BPR loss~\cite{rendle2009bpr} for LightGCN~\cite{he2020lightgcn} in the static setting and binary cross-entropy loss for SASRec~\cite{kang2018self} in the sequential setting.


When \method is trained solely with the recommendation supervision loss, we observe that the attention {weights} in the foundation model fusion process often collapse onto a single foundation model, assigning markedly higher weights to that model than to the others across all users.
To alleviate this issue, we introduce an attention-collapse regularization loss $\mathcal{L}_{reg}$ that prevents the attention {weights} from becoming overly concentrated on a single foundation model.
Specifically, the loss is computed as follows:
\begin{equation*}
\mathcal{L}_{reg}=\sum\nolimits_{p=1}^{\vert\mathcal{F}\vert}\overline{\alpha}_{p}\log{(\overline{\alpha}_{p})}, \text{ where } \overline{\alpha}_{p}=\frac{1}{\vert\mathcal{U}\vert}\sum\nolimits_{u\in\mathcal{U}} \alpha_{up}.
\end{equation*}
Note that $\mathcal{L}_{reg}$ penalizes cases where a single foundation model receives dominant attention weights across users.
Importantly, the regularization is applied to the fusion weights aggregated across users rather than to each user's weights individually. 
Thus, it discourages a single foundation model from dominating globally while still allowing individual users to place high weight on different models.\footnote{{A potential concern is that $\mathcal{L}_{\mathrm{reg}}$ could yield uniform weights across foundation models for every user. 
The loss can also reach the same minimum even when different users assign different weights to the same foundation model, as long as all foundation models receive equal weight on average across users. 
Our experiments show that the learned attention weights assigned to foundation models vary across users (Section~\ref{subsec:exp-attentionweight}).}}
The overall training objective is defined as $\mathcal{L}=\mathcal{L}_{rec}+\lambda\mathcal{L}_{reg}$, where $\lambda$ controls the contribution of the regularization loss, and all learnable parameters are jointly optimized to minimize $\mathcal{L}$.

\subsubsection{Inference}
At inference time, \method computes recommendation scores using the same forward-pass procedure as during training.
Specifically, 
after encoding item metadata using $\vert\mathcal{F}\vert$ foundation models (Section~\ref{subsec:method-dataencoding}), 
\method computes a recommendation score for each foundation model (Section~\ref{subsec:method-scoring}) 
and then performs user-adaptive fusion over these scores (Section~\ref{subsec:method-fusion}) to obtain the final recommendation score for each target user-item pair.



    \section{Experiments}
    \label{sec:experiment}
    

\begin{table*}[t]
\centering
\caption{\textit{Recommendation performance in the static setting}. 
Mean and standard deviation of Recall@$K$ (R@$K$) and NDCG@$K$ (N@$K$) over five experimental trials are reported, where $K \in \{20, 30\}$.
All values are multiplied by 100 for better readability.
The best-performing results are highlighted in \bestbox.
* indicates cases where \method statistically significantly outperforms the best-performing baseline in the corresponding setting, based on a one-sided Wilcoxon signed-rank test ($p<0.05$).}
\label{tab:main_static_results}
\renewcommand{\arraystretch}{1.0} 
\setlength{\tabcolsep}{2.0pt}
\resizebox{\textwidth}{!}{
\begin{tabular}{c | l | cccc | cccc | cccc | cccc}
\toprule
& \multirow{2}{*}{Methods}
& \multicolumn{4}{c|}{Sports \& Outdoors}
& \multicolumn{4}{c|}{Toys \& Games}
& \multicolumn{4}{c|}{Beauty \& Personal Care}
& \multicolumn{4}{c}{Arts, Crafts \& Sewing} \\
\cmidrule(lr){3-6}
\cmidrule(lr){7-10}
\cmidrule(lr){11-14}
\cmidrule(lr){15-18}
& & R@20 & R@30 & N@20 & N@30
& R@20 & R@30 & N@20 & N@30
& R@20 & R@30 & N@20 & N@30
& R@20 & R@30 & N@20 & N@30 \\

\midrule
\midrule


\multirow{4}{*}{\rotatebox{90}{Single}} & OpenAI~\cite{openai_text_embedding_3_large}
& 6.08 \std{.08} & 7.78 \std{.11} & 2.64 \std{.05} & 3.01 \std{.05}
& 6.58 \std{.08} & 8.19 \std{.12} & 2.95 \std{.05} & 3.29 \std{.06}
& 5.15 \std{.05} & 6.61 \std{.03} & 2.15 \std{.02} & 2.46 \std{.02}
& 7.65 \std{.09} & 9.77 \std{.07} & 3.26 \std{.04} & 3.71 \std{.03} \\

& Gemini~\cite{lee2025gemini}
& 5.87 \std{.09} & 7.50 \std{.12} & 2.49 \std{.04} & 2.84 \std{.05}
& 6.15 \std{.06} & 7.65 \std{.09} & 2.74 \std{.02} & 3.06 \std{.03}
& 4.61 \std{.05} & 6.02 \std{.06} & 1.90 \std{.02} & 2.21 \std{.02}
& 7.33 \std{.05} & 9.36 \std{.06} & 3.09 \std{.02} & 3.52 \std{.01} \\

& SigLIP2~\cite{tschannen2025siglip}
& 5.81 \std{.02} & 7.42 \std{.07} & 2.55 \std{.01} & 2.89 \std{.02}
& 6.57 \std{.11} & 8.18 \std{.10} & 2.95 \std{.05} & 3.30 \std{.05}
& 5.23 \std{.04} & 6.61 \std{.05} & 2.18 \std{.03} & 2.51 \std{.03}
& 7.33 \std{.08} & 9.28 \std{.09} & 3.13 \std{.03} & 3.55 \std{.02} \\

& PE~\cite{bolya2026perception}
& 5.82 \std{.06} & 7.40 \std{.09} & 2.51 \std{.03} & 2.85 \std{.03}
& 6.41 \std{.06} & 7.94 \std{.08} & 2.85 \std{.03} & 3.18 \std{.04}
& 5.19 \std{.03} & 6.55 \std{.04} & 2.18 \std{.02} & 2.47 \std{.02}
& 7.31 \std{.07} & 9.26 \std{.13} & 3.14 \std{.04} & 3.55 \std{.05} \\

\midrule

\multirow{2}{*}{\rotatebox{90}{Align}} & OpenCLIP~\cite{radford2021learning}
& 6.04 \std{.12} & 7.78 \std{.10} & 2.60 \std{.06} & 2.97 \std{.05}
& 6.47 \std{.04} & 8.00 \std{.05} & 2.92 \std{.02} & 3.24 \std{.02}
& 5.05 \std{.04} & 6.50 \std{.05} & 2.13 \std{.03} & 2.44 \std{.03}
& 7.63 \std{.10} & 9.79 \std{.11} & 3.26 \std{.07} & 3.72 \std{.08} \\

& MetaCLIP2~\cite{chuang2025meta}
& 6.15 \std{.07} & 7.84 \std{.05} & 2.65 \std{.03} & 3.01 \std{.03}
& 6.70 \std{.12} & 8.27 \std{.03} & 3.00 \std{.06} & 3.33 \std{.04}
& 5.18 \std{.04} & 6.69 \std{.04} & 2.18 \std{.02} & 2.51 \std{.02}
& 7.72 \std{.06} & 9.83 \std{.04} & 3.30 \std{.03} & 3.75 \std{.02} \\

\midrule

\multirow{2}{*}{\rotatebox{90}{Multi}} & FREEDOM~\cite{zhou2023tale}
& 6.11 \std{.10} & 7.88 \std{.08} & 2.61 \std{.05} & 2.99 \std{.05}
& 6.83 \std{.06} & 8.48 \std{.08} & 3.14 \std{.03} & 3.47 \std{.03}
& 5.13 \std{.04} & 6.56 \std{.02} & 2.17 \std{.02} & 2.48 \std{.01}
& 7.65 \std{.09} & 9.74 \std{.09} & 3.31 \std{.04} & 3.73 \std{.03} \\

& DA-MRS~\cite{xv2024improving}
& 6.19 \std{.05} & 7.99 \std{.07} & 2.77 \std{.02} & 3.15 \std{.01}
& 6.87 \std{.07} & 8.51 \std{.06} & \best 3.14 \std{.06} & 3.49 \std{.05}
& 5.13 \std{.03} & 6.44 \std{.03} & 2.20 \std{.01} & 2.48 \std{.01}
& 7.85 \std{.07} & 10.01 \std{.07} & 3.35 \std{.03} & 3.80 \std{.03} \\

\midrule

\multirow{5}{*}{\rotatebox{90}{Fusion}} & Mean
& 6.06 \std{.03} & 7.81 \std{.05} & 2.66 \std{.01} & 3.03 \std{.01}
& 6.59 \std{.07} & 8.27 \std{.07} & 2.93 \std{.03} & 3.29 \std{.03}
& 5.25 \std{.03} & 6.68 \std{.04} & 2.19 \std{.01} & 2.50 \std{.01}
& 7.73 \std{.05} & 9.75 \std{.02} & 3.27 \std{.02} & 3.70 \std{.01} \\

& Concatenation
& 6.17 \std{.08} & 7.85 \std{.05} & 2.69 \std{.05} & 3.05 \std{.04}
& 6.68 \std{.08} & 8.30 \std{.09} & 3.00 \std{.03} & 3.35 \std{.03}
& 5.26 \std{.03} & 6.68 \std{.03} & 2.22 \std{.02} & 2.53 \std{.02}
& 7.86 \std{.09} & 9.91 \std{.12} & 3.36 \std{.01} & 3.79 \std{.02} \\

& DME~\cite{kiela2018dynamic}
& 6.10 \std{.10} & 7.79 \std{.10} & 2.68 \std{.03} & 3.03 \std{.03}
& 6.66 \std{.06} & 8.36 \std{.08} & 2.99 \std{.04} & 3.35 \std{.05}
& 5.13 \std{.03} & 6.59 \std{.07} & 2.15 \std{.02} & 2.46 \std{.02}
& 7.56 \std{.12} & 9.72 \std{.11} & 3.25 \std{.04} & 3.70 \std{.04} \\

& SimMMDG~\cite{dong2023simmmdg}
& 6.22 \std{.13} & 7.90 \std{.11} & 2.72 \std{.05} & 3.07 \std{.05}
& 6.66 \std{.10} & 8.25 \std{.12} & 2.98 \std{.06} & 3.31 \std{.07}
& 5.21 \std{.07} & 6.64 \std{.09} & 2.19 \std{.05} & 2.49 \std{.05}
& 7.70 \std{.06} & 9.87 \std{.09} & 3.31 \std{.02} & 3.76 \std{.02} \\

& MoME~\cite{shen2024mome} 
& 6.31 \std{.03} & 7.99 \std{.03} & 2.73 \std{.01} & 3.08 \std{.01}
& 6.76 \std{.07} & 8.49 \std{.06} & 3.05 \std{.01} & 3.42 \std{.02} 
& 5.09 \std{.01} & 6.62 \std{.01} & 2.12 \std{.01} & 2.45 \std{.01} 
& 7.85 \std{.01} & 10.02 \std{.04} & 3.33 \std{.02} & 3.79 \std{.02} \\

\midrule
\midrule

& \method (Ours)
& \best 6.56 \std{.06}* & \best 8.40 \std{.04}* & \best 2.86 \std{.03}* & \best 3.25 \std{.03}*
& \best 6.93 \std{.06} & \best 8.76 \std{.10}* &  3.11 \std{.05} & \best 3.50 \std{.05}
& \best 5.36 \std{.05}* & \best 6.86 \std{.09}* & \best 2.24 \std{.02}* & \best2.56 \std{.02}*
& \best 8.02 \std{.05}* & \best 10.32 \std{.10}* & \best 3.42 \std{.03}* & \best 3.91 \std{.04}* \\

\bottomrule
\end{tabular}
}
\end{table*}


In this section, we evaluate the effectiveness of \method in recommendation by answering the following five questions:

\begin{itemize}[leftmargin=*]
    \item \textbf{\textit{Q1.}} How effective is \method in the static setting?

    \item \textbf{\textit{Q2.}} How effective is \method in the sequential setting?



    \item \textbf{\textit{Q3.}}  {How fast is \method compared to baselines at inference time?}

    \item \textbf{\textit{Q4.}} Does the learned user attention weight distribution align with our intuition?

    \item \textbf{\textit{Q5.}} How does each key component of \method contribute to its overall performance?
\end{itemize}

\subsection{Experimental setup}

\subsubsection{\textbf{Datasets.}}
As in our analysis (Section~\ref{sec:analysis}), we use four domain-specific e-commerce datasets released by~\citet{hou2024bridging}: (1) Sports \& Outdoors (Sports), (2) Toys \& Games (Toys), (3) Beauty \& Personal Care (Beauty), and (4) Arts, Crafts, and Sewing (Arts). 
We use the preprocessed datasets provided by~\citet{kim2026personalized}, with each item's title and image serving as its textual and visual metadata, respectively.
Further details about the datasets are provided in~\cite{anonymousgithub}.

\subsubsection{\textbf{Baseline methods and \method}}
For comparison, we consider four categories of baselines, comprising 13 methods in total. 
Below, we describe details of each category.

\begin{itemize}[leftmargin=*]
    
    \item \textbf{Baseline 1 (Single): Single-foundation-model-based methods.} 
    This baseline category allows us to assess the benefit of using multiple foundation models over a single foundation model.
    Specifically, we use the four foundation models considered in our analysis (Section~\ref{sec:analysis}): OpenAI's large text embedding model (OpenAI)~\cite{openai_text_embedding_3_large}, 
    Gemini text embedding model (Gemini)~\cite{lee2025gemini}, SigLIP2~\cite{tschannen2025siglip} and Perception Encoder (PE)~\cite{bolya2026perception}.
    These methods use the same backbone models as in Section~\ref{subsec:analysis-setup}.
    
    \item \textbf{Baseline 2 (Align): Aligned-multimodal-foundation-model-based methods.}
    This baseline category provides a comparison with several representative methods that use aligned multimodal embeddings, where representations from different modalities can be readily fused due to their shared embedding space.
    Specifically, we use two types of multimodal foundation models: OpenCLIP~\cite{radford2021learning} and MetaCLIP2~\cite{chuang2025meta}.
    In these methods, we treat the text and vision representations as separate foundation-model views, obtain a recommendation score from each view following Sections~\ref{subsec:method-dataencoding} and~\ref{subsec:method-scoring}, and average the resulting scores.

    \item \textbf{Baseline 3 (Multi): Multimodal recommender systems.}
    This baseline category provides a comparison with recommender systems that leverage multimodal item-metadata representations, using a single foundation model for each modality.
    Specifically, we use FREEDOM~\cite{zhou2023tale} and DA-MRS~\cite{xv2024improving} as static multimodal recommender systems, and M$^3$Rec~\cite{guo2025m} and DeCoRec~\cite{chen2025decorec} as sequential multimodal recommender systems.
    Since these methods use one text foundation model and one vision foundation model, we use OpenAI's large text embedding model~\cite{openai_text_embedding_3_large} and SigLIP2~\cite{tschannen2025siglip}, respectively. 
    Note that these foundation models achieve the best single-foundation-model performance within their respective text and vision categories.

    \item \textbf{Baseline 4 (Fusion): Representation-level fusion methods.}
    These baselines allow us to evaluate whether preserving foundation-model-specific recommendation views and performing fusion at the score level provides benefits over directly learning a fused representation.
    To this end, we consider two heuristic fusion methods and three learnable fusion methods.
    For the former, namely mean pooling and concatenation, representations are first projected into a common dimensional space and then combined into a single representation. 
    For the latter, we include representative methods designed to fuse representations from diverse foundation models in computer vision and natural language processing, including DME~\cite{kiela2018dynamic}, SimMMDG~\cite{dong2023simmmdg}, and MoME~\cite{shen2024mome}.
    Each method combines representations from the four foundation models described above (i.e., OpenAI, Gemini, SigLIP2, and PE) using its fusion mechanism. 
    The resulting fused representation is then treated as a single representation and used as input to the same backbone recommender systems described in Section~\ref{subsec:analysis-setup}.

    \item \textbf{Ours: \method.}
    For \method, we use the aforementioned four foundation models, comprising two text models (OpenAI and Gemini) and two vision models (SigLIP2 and PE), consistent with those used by the fusion baselines.
    Note that, although we evaluate \method using these four representative models, its formulation can accommodate different numbers of foundation models.
\end{itemize}

\subsubsection{\textbf{Training, evaluation, and hyperparameters}}
For the train/validation/test split in both static and sequential settings, we adopt the leave-one-out protocol~\cite{kang2018self, sun2019bert4rec}, using each user’s second-to-last interaction for validation, the last interaction for testing, and all preceding interactions for training.
We conduct five experimental runs with different model initializations for each model–dataset configuration.
For evaluation, we use Recall@$K$ and NDCG@$K$, where $K\in\{20,30\}$.
To assess the statistical significance of \method's performance gains, we apply an exact one-sided Wilcoxon signed-rank test against the best-performing baseline in each setting, using five runs paired by random seed~\cite{wilcoxon1945individual}.
All models are trained for 50 epochs, with each checkpoint evaluated on the validation set at the end of every epoch. 
We select the checkpoint with the highest validation Recall@30 and use it for the final evaluation on the test set.
We fix the {ID-view-score balancing hyperparameter $\beta$} and the regularization-loss balancing hyperparameter $\lambda$ to 1.
Further experimental details, including the hyperparameter search spaces, {hyperparameter sensitivity analyses}, and the hyperparameter configurations achieving the best validation performance, are provided in the supplementary materials~\cite{anonymousgithub}.

\begin{table*}[t]
\centering
\caption{\textit{Recommendation performance in the sequential setting}. 
Mean and standard deviation of Recall@$K$ (R@$K$) and NDCG@$K$ (N@$K$) over five experimental trials are reported, where $K \in \{20, 30 \}$.
All values are multiplied by 100 for better readability.
The best-performing results are highlighted in \bestbox.
* indicates cases where \method statistically significantly outperforms the best-performing baseline in the corresponding setting, based on a one-sided Wilcoxon signed-rank test ($p<0.05$).} 
\label{tab:main_sequential_results}
\renewcommand{\arraystretch}{1.0} 
\setlength{\tabcolsep}{2.0pt}
\resizebox{\textwidth}{!}{
\begin{tabular}{c | l | cccc | cccc | cccc | cccc}
\toprule
& \multirow{2}{*}{Methods}
& \multicolumn{4}{c|}{Sports \& Outdoors}
& \multicolumn{4}{c|}{Toys \& Games}
& \multicolumn{4}{c|}{Beauty \& Personal Care}
& \multicolumn{4}{c}{Arts, Crafts \& Sewing} \\
\cmidrule(lr){3-6}
\cmidrule(lr){7-10}
\cmidrule(lr){11-14}
\cmidrule(lr){15-18}
& & R@20 & R@30 & N@20 & N@30
& R@20 & R@30 & N@20 & N@30
& R@20 & R@30 & N@20 & N@30
& R@20 & R@30 & N@20 & N@30 \\

\midrule

\multirow{4}{*}{\rotatebox{90}{Single}} & OpenAI~\cite{openai_text_embedding_3_large}
& 5.29 \std{.10} & 6.72 \std{.11} & 2.18 \std{.05} & 2.48 \std{.05}
& 5.44 \std{.08} & 6.96 \std{.09} & 2.35 \std{.04} & 2.67 \std{.05}
& 4.34 \std{.06} & 5.72 \std{.04} & 1.81 \std{.03} & 2.10 \std{.03}
& 6.69 \std{.12} & 8.57 \std{.12} & 2.73 \std{.05} & 3.13 \std{.06} \\

& Gemini~\cite{lee2025gemini}
& 5.10 \std{.06} & 6.47 \std{.07} & 2.10 \std{.03} & 2.39 \std{.07}
& 4.95 \std{.10} & 6.14 \std{.12} & 2.13 \std{.07} & 2.38 \std{.08}
& 4.10 \std{.05} & 5.46 \std{.09} & 1.69 \std{.02} & 1.99 \std{.02}
& 6.00 \std{.11} & 7.84 \std{.16} & 2.42 \std{.05} & 2.81 \std{.05} \\

& SigLIP2~\cite{tschannen2025siglip}
& 4.92 \std{.07} & 6.38 \std{.06} & 2.06 \std{.04} & 2.37 \std{.05}
& 5.28 \std{.08} & 6.77 \std{.08} & 2.24 \std{.04} & 2.56 \std{.04}
& 4.17 \std{.05} & 5.45 \std{.04} & 1.67 \std{.03} & 1.94 \std{.03}
& 5.99 \std{.09} & 7.86 \std{.11} & 2.47 \std{.05} & 2.86 \std{.06} \\

& PE~\cite{bolya2026perception}
& 4.90 \std{.06} & 6.41 \std{.04} & 2.10 \std{.02} & 2.42 \std{.02}
& 5.20 \std{.10} & 6.60 \std{.13} & 2.19 \std{.04} & 2.49 \std{.05}
& 4.15 \std{.06} & 5.41 \std{.06} & 1.69 \std{.02} & 1.93 \std{.02}
& 6.11 \std{.14} & 7.92 \std{.15} & 2.55 \std{.04} & 2.93 \std{.04} \\

\midrule

\multirow{2}{*}{\rotatebox{90}{Align}} & OpenCLIP~\cite{radford2021learning}
& 4.98 \std{.10} & 6.54 \std{.12} & 2.08 \std{.03} & 2.41 \std{.03}
& 5.68 \std{.07} & 7.20 \std{.07} & 2.43 \std{.05} & 2.75 \std{.05}
& 4.32 \std{.07} & 5.62 \std{.09} & 1.77 \std{.03} & 2.05 \std{.03}
& 6.26 \std{.17} & 8.12 \std{.17} & 2.59 \std{.08} & 2.99 \std{.08} \\

& MetaCLIP2~\cite{chuang2025meta}
& 5.26 \std{.07} & 6.84 \std{.05} & 2.17 \std{.05} & 2.50 \std{.05}
& 5.88 \std{.10} & 7.61 \std{.09} & 2.53 \std{.06} & 2.89 \std{.05}
& 4.58 \std{.07} & 5.88 \std{.06} & 1.91 \std{.02} & 2.19 \std{.02}
& 6.82 \std{.04} & 8.75 \std{.04} & 2.80 \std{.03} & 3.21 \std{.04} \\

\midrule

\multirow{2}{*}{\rotatebox{90}{Multi}} & M$^3$Rec~\cite{guo2025m}
& 5.45 \std{.09} & 7.04 \std{.05} & 2.25 \std{.03} & 2.59 \std{.03}
& 5.79 \std{.11} & 7.44 \std{.13} & 2.43 \std{.07} & 2.78 \std{.07}
& 4.78 \std{.03} & 6.13 \std{.05} & 2.00 \std{.02} & 2.29 \std{.02}
& 7.10 \std{.06} & 9.02 \std{.08} & 2.95 \std{.06} & 3.38 \std{.06} \\

& DeCoRec~\cite{chen2025decorec}
& 5.24 \std{.07} & 6.77 \std{.10} & 2.22 \std{.07} & 2.55 \std{.06}
& 6.30 \std{.11} & 7.85 \std{.12} & 2.73 \std{.04} & 3.09 \std{.04}
& 4.88 \std{.03} & 6.18 \std{.03} & 2.04 \std{.02} & 2.30 \std{.02}
& \best 7.23 \std{.06} & 9.13 \std{.06} & 3.01 \std{.03} & 3.41 \std{.03} \\

\midrule

\multirow{5}{*}{\rotatebox{90}{Fusion}} & Mean
& 5.30 \std{.11} & 6.74 \std{.08} & 2.30 \std{.03} & 2.62 \std{.03}
& 6.28 \std{.13} & 7.86 \std{.13} & 2.72 \std{.06} & 3.06 \std{.06}
& 4.50 \std{.07} & 5.86 \std{.11} & 1.88 \std{.04} & 2.17 \std{.04}
& 7.16 \std{.10} & 9.00 \std{.11} & 2.96 \std{.04} & 3.35 \std{.04} \\

& Concatenation
& 5.24 \std{.05} & 6.77 \std{.07} & 2.17 \std{.04} & 2.49 \std{.04}
& 5.84 \std{.08} & 7.63 \std{.08} & 2.49 \std{.04} & 2.87 \std{.03}
& 4.43 \std{.04} & 5.83 \std{.08} & 1.84 \std{.03} & 2.14 \std{.03}
& 6.98 \std{.10} & 8.89 \std{.12} & 2.88 \std{.05} & 3.28 \std{.06} \\

& DME~\cite{kiela2018dynamic}
& 5.42 \std{.06} & 6.98 \std{.16} & 2.24 \std{.06} & 2.59 \std{.06}
& 5.50 \std{.07} & 7.16 \std{.12} & 1.73 \std{.02} & 2.02 \std{.03}
& 4.42 \std{.02} & 5.90 \std{.03} & 1.65 \std{.02} & 1.98 \std{.02}
& 6.68 \std{.14} & 8.67 \std{.11} & 2.83 \std{.07} & 3.25 \std{.06} \\

& SimMMDG~\cite{dong2023simmmdg}
& 4.96 \std{.16} & 6.50 \std{.14} & 2.04 \std{.07} & 2.36 \std{.07}
& 5.01 \std{.13} & 6.74 \std{.15} & 2.14 \std{.07} & 2.45 \std{.06}
& 4.54 \std{.06} & 5.88 \std{.06} & 1.85 \std{.03} & 2.15 \std{.03}
& 6.57 \std{.11} & 8.71 \std{.12} & 2.69 \std{.07} & 3.14 \std{.07} \\

& MoME~\cite{shen2024mome} 
& 5.19 \std{.12} & 6.78 \std{.12} & 2.18 \std{.05} & 2.51 \std{.04}
& 5.33 \std{.11} & 7.02 \std{.10} & 2.22 \std{.05} & 2.58 \std{.05}
& 4.65 \std{.05} & 6.10 \std{.06} & 1.93 \std{.03} & 2.24 \std{.03}
& 6.92 \std{.08} & 8.90 \std{.11} & 2.90 \std{.03} & 3.35 \std{.03} \\

\midrule

& \method (Ours)
& \best 5.60 \std{.04}* & \best 7.24 \std{.05}* & \best 2.34 \std{.04}* & \best 2.69 \std{.04}*
& \best 6.54 \std{.06}* & \best 8.23 \std{.08}* & \best 2.83 \std{.06}* & \best 3.19 \std{.06}*
& \best 4.94 \std{.04}* & \best 6.40 \std{.03}* & \best 2.06 \std{.02} & \best 2.36 \std{.02}*
& 7.22 \std{.06} & \best 9.21 \std{.03}* & \best 3.04 \std{.04} & \best 3.46 \std{.04} \\

\bottomrule
\end{tabular}
}
\end{table*}

\subsection{Q1. Performance in the static setting}\label{subsec:exp-static}

As shown in Table~\ref{tab:main_static_results}, \method outperforms all the baseline methods in 15 out of 16 settings, and its improvement over the strongest baseline in each setting is statistically significant in 13 out of 16 settings.
Among the baselines, DA-MRS~\cite{xv2024improving}, a multimodal recommender system, achieves the best average ranking. 
The strong performance of both \method and DA-MRS over foundation-model fusion methods adapted from NLP and CV highlights the importance of designing fusion approaches specifically for recommendation.
Nevertheless, \method achieves better performance than DA-MRS in most settings, suggesting that employing multiple foundation models within the same modality, rather than relying on a single foundation model per modality, can further improve recommendation performance.

\subsection{Q2. Performance in the sequential setting}\label{subsec:exp-sequential}

As shown in Table~\ref{tab:main_sequential_results}, \method outperforms all baseline methods in 15 out of 16 settings, and its improvement over the strongest baseline in each setting is statistically significant in 12 out of 16 settings.
Overall, the relative performance of the methods is similar to that observed in the static setting (Section~\ref{subsec:exp-static}).
In particular, DeCoRec~\cite{chen2025decorec}, a multimodal sequential recommender system, achieves the best average ranking among the baseline methods. 
Nevertheless, \method achieves better performance than DeCoRec in most settings, suggesting that the benefit of leveraging multiple foundation models within the same modality also holds in the sequential setting.

\begin{figure}[t]
    \centering
    \includegraphics[width=0.8\linewidth]{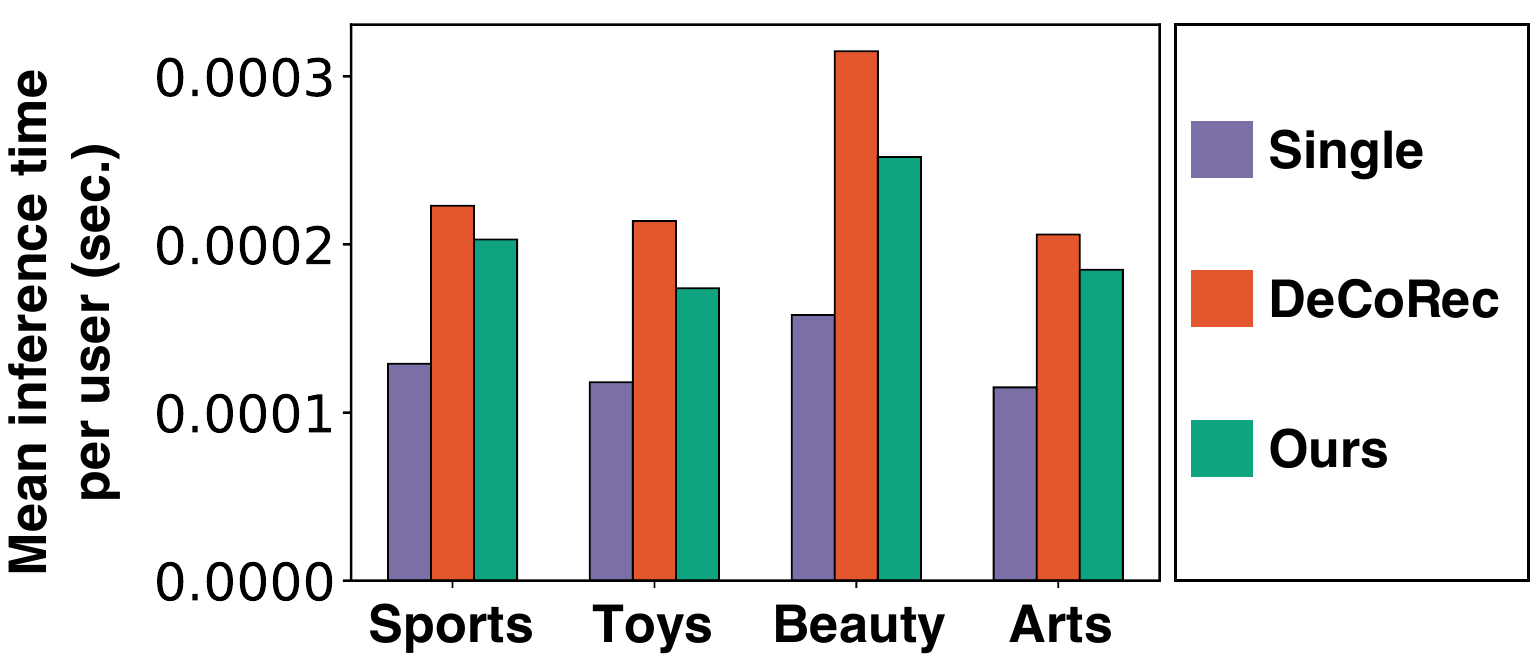}
    \caption{\textit{Inference time analysis.}
    The average inference time per user of \method is lower than that of the strongest baseline, DeCoRec~\cite{chen2025decorec}. 
    Moreover, compared with using a single foundation model (Single), \method incurs an additional inference time of less than $10^{-4}$ seconds per user, indicating that its practical overhead is minimal.}
    \label{fig:inferencetime}
\end{figure}

\subsection{Q3. Inference time analysis}\label{subsec:inferencetime}

\noindent\textbf{Setup.}
In this section, we evaluate the inference speed of \method, an important consideration for deploying recommender systems in practice.
To this end, we compare the inference speed of \method with that of (i) a single-foundation-model variant using OpenAI embeddings~\cite{openai_text_embedding_3_large} and (ii) DeCoRec~\cite{chen2025decorec}, the strongest baseline in the sequential recommendation setting.

\noindent\textbf{Result.}
The results are in Figure~\ref{fig:inferencetime}.
First, \method achieves faster inference than its strongest competitor, DeCoRec. 
Second, despite leveraging four foundation-model views, \method does not require four times the inference time of a single-foundation-model setting and instead incurs less than 60\% additional inference time on average across datasets.
{This is because (1) the foundation models remain frozen, allowing item metadata embeddings to be precomputed, and (2) recommendation scores across views can be computed in parallel.}
Moreover, the additional inference time is not substantial in practice, with an average increase of less than $10^{-4}$ seconds per user. 
These results suggest that \method introduces only modest practical overhead compared with using a single foundation model.

\begin{table*}[t]
\centering
\caption{\textit{Ablation study}. 
Mean and standard deviation of Recall@$K$ (R@$K$) and NDCG@$K$ (N@$K$) over five experimental trials are reported, where $K \in \{20,30\}$.
All values are multiplied by 100 for better readability.
The best-performing results are highlighted in \bestbox.
Overall, \method achieves the best performance in 15 out of 16 cases.}
\label{tab:main_ablation}
\renewcommand{\arraystretch}{1.05} 
\setlength{\tabcolsep}{2.0pt}
\resizebox{\textwidth}{!}{
\begin{tabular}{l | cccc | cccc | cccc | cccc}
\toprule
\multirow{2}{*}{Methods}
& \multicolumn{4}{c|}{Sports \& Outdoors}
& \multicolumn{4}{c|}{Toys \& Games}
& \multicolumn{4}{c|}{Beauty \& Personal Care}
& \multicolumn{4}{c}{Arts, Crafts \& Sewing} \\
\cmidrule(lr){2-5}
\cmidrule(lr){6-9}
\cmidrule(lr){10-13}
\cmidrule(lr){14-17}
& R@20 & R@30 & N@20 & N@30
& R@20 & R@30 & N@20 & N@30
& R@20 & R@30 & N@20 & N@30
& R@20 & R@30 & N@20 & N@30 \\

\midrule
\midrule


w/o Text
& 5.00 \std{.10} & 6.58 \std{.11} & 2.10 \std{.05} & 2.43 \std{.04}
& 5.82 \std{.11} & 7.38 \std{.12} & 2.49 \std{.04} & 2.82 \std{.04}
& 4.16 \std{.10} & 5.50 \std{.09} & 1.72 \std{.04} & 2.00 \std{.03}
& 6.33 \std{.07} & 8.18 \std{.06} & 2.63 \std{.02} & 3.02 \std{.03} \\

w/o Vision
& 5.30 \std{.12} & 6.93 \std{.12} & 2.19 \std{.06} & 2.53 \std{.06}
& 5.31 \std{.08} & 6.82 \std{.10} & 2.28 \std{.06} & 2.60 \std{.06}
& 4.48 \std{.07} & 5.85 \std{.07} & 1.84 \std{.04} & 2.13 \std{.04}
& 6.72 \std{.14} & 8.68 \std{.15} & 2.81 \std{.08} & 3.22 \std{.08} \\

w/o Reg
& 5.52 \std{.09} & 7.15 \std{.09} & 2.25 \std{.03} & 2.61 \std{.02}
& 6.09 \std{.13} & 7.78 \std{.15} & 2.66 \std{.06} & 3.02 \std{.05}
& 4.64 \std{.06} & 6.09 \std{.06} & 2.00 \std{.03} & 2.31 \std{.04}
& 7.06 \std{.11} & 9.04 \std{.10} & 2.96 \std{.04} & 3.38 \std{.04} \\

1-FM-Per-Mod
& 5.28 \std{.07} & 6.88 \std{.08} & 2.21 \std{.04} & 2.55 \std{.04}
& 5.81 \std{.08} & 7.50 \std{.10} & 2.47 \std{.03} & 2.83 \std{.04}
& 4.44 \std{.05} & 5.86 \std{.05} & 1.83 \std{.03} & 2.13 \std{.03}
& 6.82 \std{.08} & 8.81 \std{.08} & 2.88 \std{.04} & 3.30 \std{.05} \\

Mean-pool
& 5.28 \std{.08} & 7.12 \std{.09} & 2.20 \std{.05} & 2.52 \std{.05}
& 6.41 \std{.09} & 8.00 \std{.11} & 2.65 \std{.04} & 2.99 \std{.04}
& 4.62 \std{.06} & 6.01 \std{.05} & 1.90 \std{.02} & 2.21 \std{.02}
& 7.05 \std{.07} & 9.11 \std{.07} & 3.00 \std{.03} & 3.41 \std{.04} \\

Global-pool
& 5.24 \std{.05} & 7.12 \std{.05} & 2.18 \std{.03} & 2.57 \std{.04}
& 6.42 \std{.08} & 8.03 \std{.09} & 2.65 \std{.05} & 3.00 \std{.05}
& 4.62 \std{.04} & 5.98 \std{.04} & 1.91 \std{.03} & 2.22 \std{.02}
& 7.11 \std{.08} & \best 9.27 \std{.07} & 3.00 \std{.03} & 3.42 \std{.03} \\

Item-pool
& 5.55 \std{.06} & 7.15 \std{.08} & 2.30 \std{.03} & 2.64 \std{.03}
& 6.03 \std{.11} & 7.71 \std{.10} & 2.60 \std{.07} & 2.96 \std{.07}
& 4.48 \std{.10} & 5.92 \std{.08} & 1.87 \std{.04} & 2.17 \std{.03}
& 6.88 \std{.09} & 8.88 \std{.09} & 2.79 \std{.04} & 3.22 \std{.04} \\

\midrule

\method (Ours)
& \best 5.60 \std{.04} & \best 7.24 \std{.05} & \best 2.34 \std{.04} & \best 2.69 \std{.04}
& \best 6.54 \std{.06} & \best 8.23 \std{.08} & \best 2.83 \std{.06} & \best 3.19 \std{.06}
& \best 4.94 \std{.04} & \best 6.40 \std{.03} & \best 2.06 \std{.02} & \best 2.36 \std{.02}
& \best 7.22 \std{.06} & 9.21 \std{.03} & \best 3.04 \std{.04} & \best 3.46 \std{.04} \\

\bottomrule
\end{tabular}
}
\end{table*}

\subsection{Q4. User attention weight analysis}\label{subsec:exp-attentionweight}

\noindent\textbf{Setup.}
A key design objective of our fusion strategy is to enable \method to adaptively determine the contribution of each foundation model for each user. 
Accordingly, we expect the learned attention {weights} to vary across users and foundation-model views, rather than remaining nearly uniform or globally collapsing onto a single foundation model.
To examine whether this behavior emerges in practice, we analyze the learned attention {weights} for each foundation model across users. 
Specifically, for each foundation model $f^{(p)}\in\mathcal{F}$, we collect its user-specific attention {weights} $\alpha_{up}$ (see Eq.~\eqref{eq:attention}) for all $u\in\mathcal{U}$ and visualize their distribution using a boxplot.
We evaluate this behavior under the static setting.

\begin{figure}[t]
    \centering
    \includegraphics[width=0.9\linewidth]{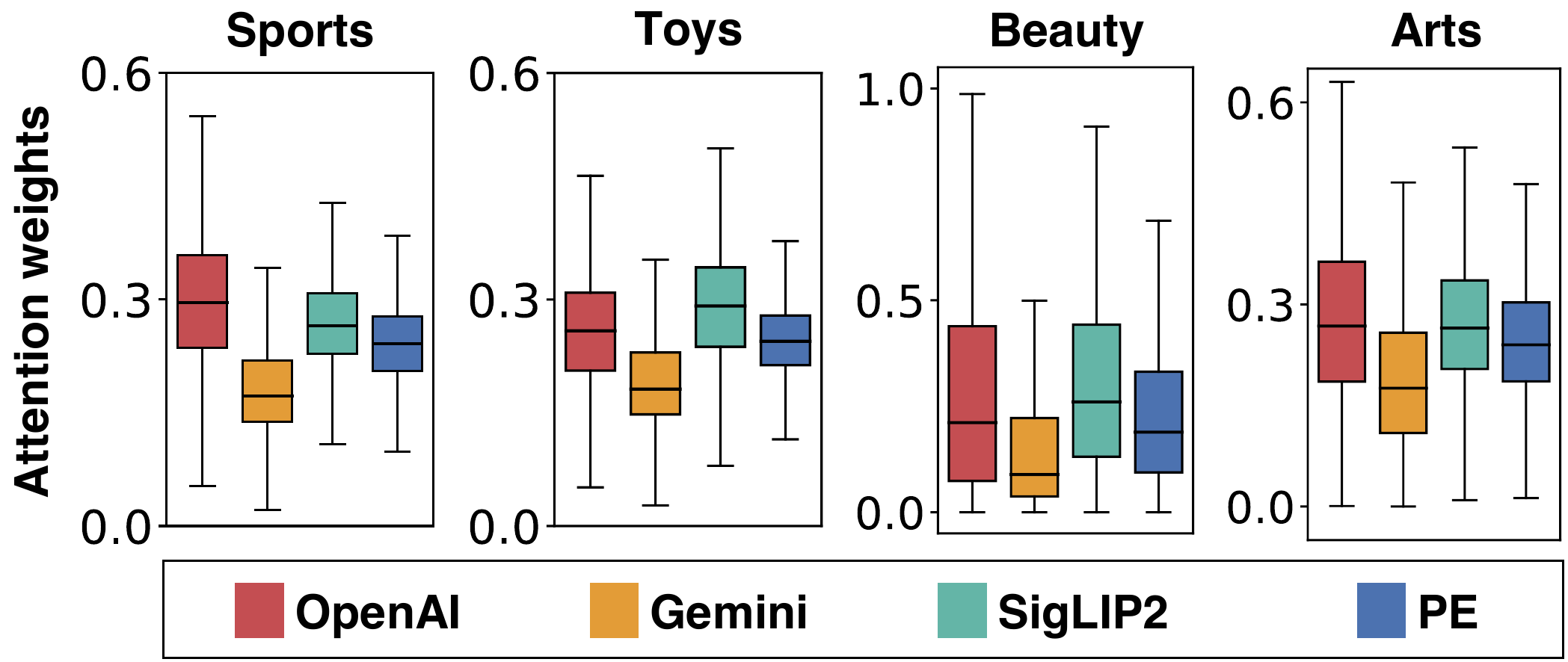}
    \caption{\textit{Distribution of learned user attention {weights}.}
    The learned attention weights exhibit substantial variation across users and foundation models, suggesting nontrivial view-weighting behavior.}
    \label{fig:designobjective}
\end{figure}

\noindent\textbf{Result.}
As shown in Figure~\ref{fig:designobjective}, the learned user attention {weights} exhibit variation across users, with some values approaching 0.9. 
The distributions are neither concentrated around the uniform weight (i.e., 0.25 in our setting) nor globally dominated by a single foundation model, indicating that \method learns heterogeneous attention patterns across users and foundation-model views. 
In addition, our supplementary materials~\cite{anonymousgithub} provide two further analyses showing that the learned attention {weights} are meaningful.
First, the performance of \method decreases when the learned attention {weights} are shuffled across users after training, suggesting that the weights are meaningfully specific to their corresponding users. 
Second, \method tends to assign higher attention {weights} to foundation models that benefit a user than to those that do not.

\subsection{Q5. Ablation study}

\noindent\textbf{Setup.}
We analyze whether the key components within \method are necessary for its strong performance. 
To this end, we consider seven variants of \method: 
(1) w/o Text, which excludes text foundation models; 
(2) w/o Vision, which excludes vision foundation models; 
(3) w/o Reg, which removes the regularization loss described in Section~\ref{subsec:method-training};
(4) 1-FM-Per-Mod, which uses one foundation model per modality: OpenAI for text and SigLIP2 for vision;
(5) Mean-pool, which replaces user-adaptive fusion with uniform averaging across foundation models;
(6) Global-pool, which replaces user-adaptive fusion with learnable global fusion weights shared across all users; and
(7) Item-pool, which replaces user-adaptive fusion with item-adaptive fusion.
We evaluate these variants in the sequential recommendation setting, following the experimental setup described in Section~\ref{subsec:exp-sequential}.

\noindent\textbf{Result.}
As shown in Table~\ref{tab:main_ablation}, \method outperforms all its variants in 15 out of 16 settings.
This result suggests that the key components of \method collectively contribute to its strong recommendation performance: leveraging both text and image foundation models, user-adaptive fusion, and a regularization loss that prevents the fusion weights from collapsing onto a single foundation model.
    
    \section{Conclusion and discussion}
    \label{sec:conclusion}

\noindent{\textbf{Conclusion.}} 
In this work, we explore how to effectively fuse multiple foundation models for recommender systems. 
To this end, we first conduct an analysis showing that the users who benefit from foundation models vary across models. 
Motivated by this observation, we introduce \method, a user-adaptive score-level fusion approach for multiple foundation models, together with a training strategy that employs a specialized regularization loss to stabilize training. 
Through extensive experiments, we demonstrate that \method outperforms existing foundation-model-fusion baselines and multimodal recommender system baselines in most cases across both static and sequential recommendation settings.

\noindent{\textbf{Future work.}}
Our analysis shows that the benefits of different foundation models vary across users. 
These differences may stem from model architecture or training data, but some models disclose limited information about these factors. 
Analyzing models with better-documented architectures and training datasets could help explain why different users benefit from different models.

    \noindent\textbf{Acknowledgements.}
    This work was partly supported by the National Research Foundation of Korea (NRF) grant funded by the Korea government (MSIT) (No. RS-2024-00406985, 30\%).
    This work was partly supported by Institute of Information \& Communications Technology Planning \& Evaluation (IITP) grant funded by the Korea government (MSIT) (No. RS-2022-II220871, Development of AI Autonomy and Knowledge Enhancement for AI Agent Collaboration, 30\%)
    (No. RS-2024-00457882, National AI Research Lab Project, 30\%)
    (No. RS-2019-II190075, Artificial Intelligence Graduate School Program (KAIST), 10\%).

    \section*{Ethical considerations}
    We believe that our study raises no major ethical concerns, as it does not involve direct human-subject experiments (e.g., crowdsourcing), and all experiments are conducted using anonymized, publicly available datasets that permit academic use. 
    Also, all foundation models used in this study permit academic use.
    Moreover, the implementations of all baseline methods are publicly available from their original providers and are permitted for academic use.

    \bibliographystyle{ACM-Reference-Format}
    \balance
	\bibliography{000Ref}
\end{document}